%% file: main.tex
\documentclass{article}

\usepackage{arxiv}

\usepackage[utf8]{inputenc} 
\usepackage[T1]{fontenc}    
\usepackage{hyperref}       
\usepackage{url}            
\usepackage{booktabs}       
\usepackage{amsfonts}       
\usepackage{nicefrac}       
\usepackage{microtype}      
\usepackage{lipsum}		
\usepackage{graphicx}
\usepackage{natbib}
\usepackage{doi}

\usepackage{multirow} 
\usepackage{graphicx} 
\usepackage{blindtext}
\usepackage{booktabs}

\usepackage{amsmath}       

\usepackage{amssymb}       

\input{shorts.tex}

\title{Faster Inference of LLMs using FP8 on the Intel Gaudi}

\author{ 
{Joonhyung Lee$^{1*}$}
	\And
{Shmulik Markovich-Golan$^{2*}$}
	\And
{Daniel Ohayon$^{2}$}
	\And
{Yair Hanani$^{2}$}
	\And
{Gunho Park$^{1}$} 
    \And
{Byeongwook Kim$^{1}$} 
    \And
{Asaf Karnieli$^{2}$}
	\And
{Uri Livne$^{2}$}
	\And
{Haihao Shen$^{2}$}
	\And
{Tai Huang$^{2}$}
    \And
{Se Jung Kwon$^{1}$}
    \And
{Dongsoo Lee$^{1}$}
    \And
}
\date{$^{1}$ NAVER Cloud, $^{2}$ Intel Corporation/Habana}

\renewcommand{\headeright}{Technical Report}
\renewcommand{\undertitle}{Technical Report}

\begin{document}
\maketitle

\input{00_abstract}


\input{01_introduction}

\input{02_background}

\input{03_implementation}

\input{04_model_evaluations}

\input{05_discussion}

\bibliographystyle{unsrtnat}
\bibliography{references}

\end{document}

%% file: shorts.tex
\newcommand{\vWT}{\mathbf{W}^{T}}
\newcommand{\vSc}{\mathbf{S}_{c}}
\newcommand{\viSc}{\mathbf{S}_{c}^{-1}}
\newcommand{\vSw}{\mathbf{S}_{w}}
\newcommand{\viSw}{\mathbf{S}_{w}^{-1}}
\newcommand{\vSx}{\mathbf{S}_{x}}
\newcommand{\viSx}{\mathbf{S}_{x}^{-1}}
\newcommand{\diag}{\textrm{diag}}
\newcommand{\vsx}{\mathbf{s}_{x}}
\newcommand{\vsw}{\mathbf{s}_{w}}
\newcommand{\vsc}{\mathbf{s}_{c}}
\newcommand{\visx}{\mathbf{s}^{-1}_{x}}
\newcommand{\visw}{\mathbf{s}^{-1}_{w}}
\newcommand{\visc}{\mathbf{s}^{-1}_{c}}
\newcommand{\vhWT}{\hat{\mathbf{W}}^{T}}
\newcommand{\rx}{r_x}
\newcommand{\sx}{s_x}
\newcommand{\rw}{r_w}
\newcommand{\sw}{s_w}
\newcommand{\tsw}{\bar{s}_w}
\newcommand{\vtsw}{\bar{\mathbf{s}}_{w}}
\newcommand{\vitSw}{\bar{\mathbf{S}}_{w}^{-1}}
\newcommand{\vtWT}{\tilde{\mathbf{W}}^{T}}
\newcommand{\vhWsT}{\hat{\mathbf{W}}_{s}^{T}}
\newcommand{\vek}{\mathbf{e}_k}

%% file: 00_abstract.tex
\begin{abstract}

Low-precision data types are essential in modern neural networks during both training and inference as they enhance throughput and computational capacity by better exploiting available hardware resources. Despite the incorporation of FP8 in commercially available neural network accelerators, a comprehensive exposition of its underlying mechanisms, along with rigorous performance and accuracy evaluations, is still lacking. In this work, we contribute in three significant ways. First, we analyze the implementation details and quantization options associated with FP8 for inference on the Intel Gaudi AI accelerator. Second, we empirically quantify the throughput improvements afforded by the use of FP8 at both the operator level and in end-to-end scenarios. Third, we assess the accuracy impact of various FP8 quantization methods. Our experimental results indicate that the Intel Gaudi 2 accelerator consistently achieves high computational unit utilization, frequently exceeding 90\% MFU, while incurring an accuracy degradation of less than 1\%.

\end{abstract}

%% file: 01_introduction.tex
\section{Introduction}\label{sec:intro}

Recent advances in AI hardware accelerators, including the NVIDIA H100, AMD MI300X, and Intel® Gaudi® 2 \& 3 AI accelerators, have facilitated the adoption of the FP8 numerical data type \citep{micikevicius2022fp8formatsdeeplearning}. FP8 offers distinct advantages over conventional 16-bit floating-point formats. In particular, FP8 reduces memory requirements by half, thereby enhancing capacity utilization and communication bandwidth, and it frequently delivers up to twice the computational throughput of 16-bit GEMM operations. Given the substantial costs associated with training and deploying large language models (LLMs), there has been increasing interest in applying FP8 formats to them \citep{peng2023fp8lmtrainingfp8large, fishman2025scaling, lee2024fp8againquantifyingeffects, kim2025investigationfp8acceleratorsllm}.

However, the reduced numerical range of FP8 can lead to potential issues when applied without consideration of numerical issues. Naïve implementations may result in overflow, where large absolute values are clipped to the maximum or minimum representable limits, or underflow, where small absolute values are rounded to zero, thereby significantly degrading accuracy. To mitigate these challenges, \citet{micikevicius2022fp8formatsdeeplearning} introduced a scaled matrix multiplication algorithm. Unlike standard 16-bit or 32-bit floating-point representations, FP8 is more suitably treated as a low bit-width group quantization scheme that employs per-group quantization. Nevertheless, the specific implementation details of this approach remain undefined, leaving the determination of the scaling mechanism and its granularity to hardware and software vendors.

Previous research on neural network quantization \citep{kwon-etal-2022-alphatuning, grattafiori2024llama3herdmodels} has indicated that employing finer-grained quantization can mitigate accuracy loss, albeit at the expense of throughput. It is important to note that overly fine granularity will eventually negate the performance benefits of FP8 compared to simply using BF16.

In this paper, we present in detail the FP8 implementation in the Intel Gaudi 2 and 3 accelerators, with a focus on the inference stage. Our study serves as a comprehensive reference for the application of FP8 in Intel Gaudi 2 and 3 accelerators and for future analysis of FP8 quantization. We explore the various supported FP8 configurations and report their empirical throughput measurements. Furthermore, we evaluate inference accuracy across a range of tasks, including common sense reasoning and MMLU \citep{hendrycks2021measuring}, for multiple LLMs using different FP8 configurations.

Our contributions in this work are as follows:
\begin{enumerate}
    \item We provide a detailed description of scaled FP8 matrix multiplication and its various configurations in the Intel Gaudi 2 and 3 accelerators.
    \item We report throughput measurements at both the operator level and end-to-end in LLMs for the prefill and decode stages, demonstrating that the Gaudi 2 accelerator achieves excellent computational unit utilization, often exceeding 90\% MFU.
    \item We evaluate the inference accuracy of various FP8 configurations in end-to-end LLM assessments.
\end{enumerate}

%% file: 02_background.tex
\section{Background}\label{sec:back}

In this section, we discuss the FP8 format as specified in \citet{micikevicius2022fp8formatsdeeplearning}. We focus on the E4M3 and E5M2 data types, where \emph{E} and \emph{M} denote the number of exponent and mantissa bits, respectively. For a fixed bit width, there is an inherent trade-off between dynamic range, determined primarily by the number of exponent bits, and precision, which is governed by the number of mantissa bits. In the FP8 context, the current best practice is to employ E4M3 for the forward pass to maximize accuracy, and use E5M2 in the backward pass during training because gradients typically exhibit a larger dynamic range. However, works such as \citet{deepseekai2024deepseekv3technicalreport} have proposed alternative methods such as using finer scaling factors and using E4M3 for both the forward and backward passes. Determining an appropriate scaling factor is important in scaled FP8 matrix multiplication because it influences the likelihood of overflow, underflow, and rounding errors in the quantized representation.

We adopt the following terminology:
\begin{itemize}
    \item \textbf{Offline vs. online quantization:} Offline quantization is performed prior to model execution while online quantization occurs during model execution.
    \item \textbf{Static vs. dynamic scaling:} Static scaling means that scaling factors are computed in advance using a calibration dataset. Dynamic scaling computes these factors in real time using statistics from current or recent samples. In both cases, online quantization is implied.
\end{itemize}

\subsection{Training, inference, and weight-only quantization}

In weight-only quantization (WoQ) for inference, matrix multiplication is executed in high precision. Prior to this operation, the weights are quantized offline, reducing storage and communication overhead. Weight-only quantization compresses and subsequently decompresses the weights online to high precision, and can be applied independently of the data types supported by hardware-accelerated matrix multiplication.

When an accelerator supports FP8 matrix multiplication, the GEMM throughput can be improved, similarly to INT8 quantization \citep{pmlr-v202-xiao23c}, in accelerators which support GEMM in integer formats. During inference, the weights remain fixed and are quantized offline, while activations must be quantized online. It is noteworthy that the outputs of the matrix multiplication are typically not maintained in FP8 in order to maintain high-accuracy, since the additional throughput improvement is marginal and most nonlinear activation functions cannot be effectively computed in this format.

\subsection{Scaling factor granularity}

The granularity of a scaling factor is determined by the size of the group that shares a common scaling parameter. The most elementary approach is per-tensor scaling. This method applies a single scaling factor across the entire tensor. Finer-grained scaling may reduce quantization error and has led to the development of alternative methods. It is possible to use different granularities for activations and weights, and the scaling strategy may vary across different layers.

A commonly adopted method is per-channel scaling, also referred to as row-wise scaling, in which each channel of a tensor is assigned its own scaling factor. When the number of elements per channel is large, the corresponding dynamic range may increase and using a single scaling factor per-channel may result in a significant quantization error. An alternative approach is to partition each channel into blocks and assign a distinct scaling factor to each block. The additional overhead resulting from finer granularity can be reduced if the blocks are aligned with the hardware-supported matrix multiplication tiles. The concept of micro-scaling floating points \citep{rouhani2023microscalingdataformatsdeep} divides tensors into blocks of 32 elements that share a scaling factor. However, this method is not currently supported by Gaudi 2 and 3 accelerators and is therefore beyond the scope of this work.

\subsection{Scaling factor dynamics}

Various approaches exist for determining scaling factors. Some of these approaches are applicable to inference, while others are better suited for training.

\subsubsection{Static scaling}

The static scaling strategy computes scaling factors from statistics obtained from a calibration dataset that reflects the typical input distribution during inference. This approach allows the scaling factors to be computed offline. It is unsuitable for training because the optimal scaling factors for activations and weights change as the model adapts. In addition, static scaling may incur significant quantization errors for inputs that fall outside the calibration distribution. 

Static scaling also does not support per-sample scaling of activations for the following two reasons. First, per-sample scaling would require a fixed number of input samples.
Second, even if per-token scaling factors were specified, there would be no information available to assign tokens to their optimal scaling factors during runtime.

\subsubsection{Just-in-time (dynamic) scaling}

The just-in-time (JiT) or dynamic scaling strategy calculates scaling factors by measuring the statistics of the current input data and applying the resulting factors during the GEMM operation. Previous work \citep{peng2023fp8lmtrainingfp8large} observed that JiT scaling can be inefficient because it requires multiple passes through memory. This is especially true for the per-tensor case where the tensor may not entirely reside in cache memory. Efficiency is improved for finer granularities, such as per-sample scaling, that fit within the cache. In such cases, inputs may be quantized to FP8 based on measured statistics with only a single access to global memory per input sample (token), thereby minimizing the overhead. JiT scaling has the advantage of using real-time input statistics to achieve high quantization accuracy and is applicable to both training and inference. One drawback is that tensor parallelism during inference may lead to different scaling factors across tensors when per-channel quantization, which can result in slightly different outputs across different tensor parallelism configurations.

\subsubsection{Delayed scaling}

The delayed scaling strategy, which is designed for FP8 training, computes scaling factors from previous inputs by maintaining a moving history of statistics. The scaling factors computed from this history are used for the current input. This method enables the scaling factor to be determined independently of the current sample and be computed in advance. It can be regarded as a modified version of static scaling. By recursively updating the calibration dataset with recent inputs, delayed scaling becomes suitable for training while alleviating the latency typically associated with dynamic scaling. However, it is still vulnerable to poor quantization if out-of-distribution activations emerge during training. It is unsuitable for inference as weights are fixed and activation statistics can be calculated offline on a calibration set.

\subsection{Special features of the Gaudi Accelerators}

The Gaudi accelerators provide hardware-supported features designed for scaled FP8 matrix multiplication. Key features include:
\begin{itemize}
    \item \textbf{Power-of-two scaling:} When both input tensors use per-tensor scaling with factors that are powers of two, the scaling can be efficiently implemented by adjusting the exponent bias instead of multiplying individual elements. This optimization can improve FP8 throughput by several percentage points. Table \ref{tab:fp8-gemm} demonstrates this effect. It is important to note that if only one of the input tensors uses a power-of-two scaling factor, the throughput improvement is reduced. Also, it can only be applied to per-tensor scaling.
    \item \textbf{Stochastic rounding:} During the casting operation from high-precision floating point to FP8, stochastic rounding may be applied. This unbiased rounding method introduces increased quantization noise but is potentially beneficial during training, where bias can have a negative effect. The overhead associated with including stochastic rounding is negligible compared to standard FP8 casting with rounding to the nearest value. In this context, stochastic rounding is neither required nor supported in the accumulator used in matrix multiplication since it is computed here in high-precision. Other approaches \citep{doi:10.1137/22M1510819}, which use lower precision for the accumulation, apply it to avoid bias.
\end{itemize}

The Gaudi 3 shows notable enhancements over the Gaudi 2 with respect to these FP8 features:
\begin{itemize}
    \item In the Gaudi 2, the implementation of E4M3 follows the IEEE standard for floating-point arithmetic. The largest exponent is reserved for NaN (not-a-number) and infinity. This limits the range to $\pm 240$. In the Gaudi 3, the maximal exponent is available for normal numbers. This extends the numerical range of E4M3 to $\pm 448$ as per \citet{micikevicius2022fp8formatsdeeplearning}.
    \item In terms of hardware-accelerated power-of-two scaling factors, the Gaudi 3 supports any scaling factor within the range $[2^{-32}, 2^{-31}, \dots, 2^{0}, \dots, 2^{30}, 2^{31}]$. The Gaudi 2 is limited to scaling factors in the set $[2^{-8},2^{-4}, 2^{0}, 2^{4}]$.
\end{itemize}

%% file: 03_implementation.tex
\section{Model Quantization}\label{sec:imp}
In this section, we describe the procedure for post-training quantization of a model and its subsequent deployment on Gaudi hardware. The discussion is limited to the quantization of linear operations. Other operations, e.g., \textit{softmax} and activation functions, are not considered.

Consider the quantization of the $l$-th linear layer. In its high-precision form, the computation is expressed as:
\begin{align}
\mathbf{X}_{l+1} = \mathbf{X}_{l}\vWT,
\end{align}
where $\mathbf{X}_{l}$ and $\mathbf{X}_{l+1}$ denote the input and output activations and $\mathbf{W}$ is the weight matrix with dimensions $N_{l}\times C_{l}$, $N_{l}\times C_{l+1}$, and $C_{l+1}\times C_{l}$, respectively.

The corresponding quantized version is formulated as:
\begin{align}
\mathbf{X}_{l+1}=\vSx\left(\hat{\mathbf{X}}_{l,s}\otimes \hat{\mathbf{W}}_s^{T}\right)\vSw
\end{align}
with
\begin{subequations}
\begin{align}
\hat{\mathbf{X}}_{l,s}=&\ Q\left(\mathbf{X}_{l,s}\right)\label{eq:vXs}\\
\hat{\mathbf{W}}_s^{T}=&\ Q\left(\mathbf{W}_s^{T}\right)\label{eq:vWTs}
\end{align}
\end{subequations}
and
\begin{subequations}
\begin{align}
\mathbf{X}_{l,s}=&\ \viSx\mathbf{X}_{l}\viSc\\
\mathbf{W}_s^{T}=&\ \vSc\vWT\viSw
\end{align}
\end{subequations}

Here, $\mathbf{X}_{l,s}$ and $\mathbf{W}_s^{T}$ represent the scaled versions of the activation and weight matrices, and $Q(\cdot)$ denotes the quantization operation. The operator $\otimes$ indicates a mixed-precision matrix multiplication where low-precision inputs are multiplied with the accumulation performed in FP32.

The diagonal scale matrices
\begin{subequations}
\begin{align}
    \vSx =&\  \diag\left(\vsx\right)\\
    \vSw =&\  \diag\left(\vsw\right)\\
    \vSc =&\  \diag\left(\vsc\right)
\end{align}
\end{subequations}
correspond to the input, weight, and common-dimension scales. The vectors $\vsx$, $\vsw$, and $\vsc$ are placed on the diagonals of these matrices, which have dimensions $N_{l}\times N_{l}$, $C_{l+1}\times C_{l+1}$, and $C_{l}\times C_{l}$. This formulation is sufficiently general to accommodate various scaling methods, as discussed in Sec.~\ref{subsec:scale}. Typically, the scales are selected such that the dynamic range of the scaled tensors aligns with the full range of the quantized data type, denoted by $r_q$, without causing clipping. Note that, due to the diagonal structure of the scale matrices, scaling is performed element-wise rather than by matrix multiplication.

For instance, the operation $\mathbf{X}_{l}\viSc$ is equivalent to 
\begin{subequations}
\begin{align}
\mathbf{X}_{l}\odot(\mathbf{1}_{N_{l}\times 1}\visc)
\end{align}
\end{subequations}

where
\begin{subequations}
\begin{align}
    \visx =& \left[s^{-1}_{x,1}, \ldots, s^{-1}_{x,N_{l}}\right]^{T}\\
    \visw =& \left[s^{-1}_{w,1}, \ldots, s^{-1}_{w,C_{l+1}}\right]^{T}\\
    \visc =& \left[s^{-1}_{c,1}, \ldots, s^{-1}_{c,C_{l}}\right]^{T}
\end{align}
\end{subequations}
Furthermore, since this analysis concerns the inference stage, the model weights are static and can be quantized offline.

Regarding the scale calculations, we distinguish between \emph{static} and \emph{dynamic} scaling of activations, where the scales are computed \emph{offline} and \emph{online}, respectively. For static scaling, a calibration phase is necessary to measure tensor statistics over a limited calibration dataset (see Sec.~\ref{subsec:calib}). Various methods for calculating scales are described in Sec.~\ref{subsec:scale} and a \emph{recipe} for quantizing a model is given in Sec.~\ref{subsec:recipe}. Note that for both \emph{static} and \emph{dynamic} scaling, activations are quantized \emph{online} and weights are quantized \emph{offline}, regardless of how the scaling factors for the activations are obtained.

\begin{figure}
    \centering
    \includegraphics[width=0.7\linewidth]{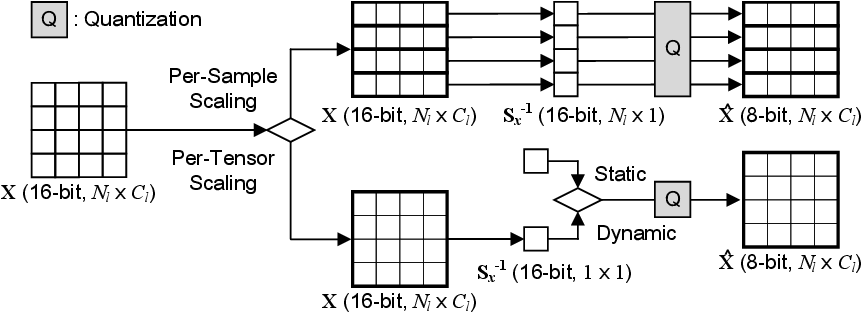}
    \caption{Activation quantization block-diagram. The activation undergoes either per-sample or per-tensor scaling. Per-sample scaling is impractical for static scaling of activations because per-token information is unknown during calibration. Per-sample scaling would also require a fixed number of samples.}
    \label{fig:act_quant_block_diagram}
\end{figure}
\begin{figure}
    \centering
    \includegraphics[width=0.8\linewidth]{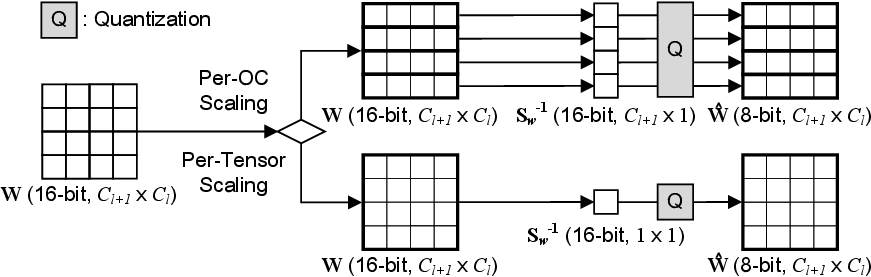}
    \caption{Weight quantization block-diagram. The high-precision weight undergoes either per-output-channel or per-tensor scaling. Weight quantization is static for inference.}
    \label{fig:weight_quant_block_diagram}
\end{figure}

\begin{figure}
    \centering
    \includegraphics[width=0.5\linewidth]{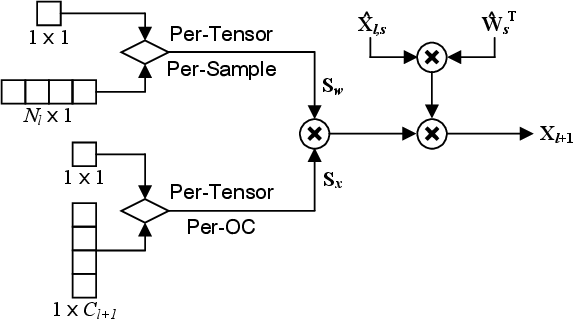}
    \caption{Descaling of the scaled FP8 GEMM operation. Scaling factors are multiplied with one another, then applied to the GEMM results.}
    \label{fig:dequant_block_diagram}
\end{figure}

\subsection{Calibration}\label{subsec:calib}

In the calibration stage, typical inputs that exhibit statistical properties similar to those expected during inference are processed and the activation statistics are measured. The measured statistics may consist of minimum and maximum values, maximum absolute value, average absolute value, or a histogram. The statistics are computed on a per-tensor, per-channel, or per-sample basis. In this work, we measure the per-tensor and per-channel maximum absolute value statistics. These are defined as follows:

\begin{subequations}
\begin{align}
r_{x} =& \max_{i,n,c}  |X^i_{l,nc}| \label{eq:act_maxabs_pt}\\
\mathbf{r}_{x|} =& \left[\max_{i,n}  |X^i_{l,n1}|, \ldots,  \max_{i,n}  |X^i_{l,nC_l}|\right]^{T}\label{eq:act_maxabs_pc}
\end{align}
\end{subequations}
with $i$, $n$ and $c$ denoting the indices of batch, sample and channel, respectively.

When activations are dynamically scaled, the activation statistics is defined for each batch $i$, either per-tensor or per-sample, as:

\begin{subequations}
\begin{align}
r^{i}_{x} &= \max_{n,c}  |X^i_{l,nc}| \label{eq:act_maxabs_pt_dynamic}\\
\mathbf{r}^{i}_{x-} =& \left[\max_{c}  |X^i_{l,1c}|, \ldots,  \max_{c}  |X^i_{l,N_lc}|\right]^{T}\label{eq:act_maxabs_per_sample_dynamic_vec}\\
\end{align}
\end{subequations}

Since the weights are static, their statistics do not depend on the input. The weight statistics are defined as follows:

\begin{subequations}
\begin{align}
r_{w} =& \max_{kc}  |W_{kc}| \label{eq:w_maxabs_pt}\\
\mathbf{r}_{w-} =& \left[\max_{c}  |W_{1c}|, \ldots,  \max_{c}  |W_{C_{l+1}c}|\right]^{T}\label{eq:w_maxabs_poc}\\
\mathbf{r}_{w|} =& \left[\max_{k}  |W_{k1}|, \ldots,  \max_{k}  |W_{kC_l}|\right]^{T}\label{eq:w_maxabs_pic}
\end{align}
\end{subequations}

The above definitions correspond to the per-tensor, per-output-channel, and per-input-channel statistics, respectively.

\subsection{Scaling methods}\label{subsec:scale}

Given the statistics of the weights and activations, obtained either through calibration or computed dynamically, several scaling methods can be employed using different combinations for the scaling of weights and activations.

In most cases considered here, the scales for weights and activations are determined independently, except for the SmoothQuant \citep{xiao2024smoothquantaccurateefficientposttraining} method. Activation scales may be computed either on a per-tensor basis (see Sec.~\ref{subsec:act_scale_per_tensor}) or on a per-sample basis (see Sec.~\ref{subsec:act_scale_per_sample}). Similarly, weight scales may be determined on a per-tensor basis (see Sec.~\ref{subsec:weight_scale_per_tensor}) or on a per-output-channel basis (see Sec.~\ref{subsec:weight_scale_per_output_channel}).

An alternative approach for quantizing weights is to optimize the squared Frobenius norm of the error:
\begin{align}
    \tilde{\sigma}_{W}^2=\|\vtWT\|^2_{\textrm{Fro}}
\end{align}
where the quantization error is defined as
\begin{align}
    \vtWT=\vhWT-\vWT
\end{align}
and
\begin{align}
    \vhWT = \viSc \vhWsT \vSw
\end{align}
For further details on scale optimization for weights using per-tensor scaling and per-channel scaling, refer to Sec.~\ref{subsec:weight_opt_pts} and Sec.~\ref{subsec:weight_opt_pcs}.

Alternatively, one may jointly determine the scales of activations and weights on a per-channel basis (see Sec.~\ref{subsec:scale_smooth_quant}). A common practice is to round the scales to a power of 2, i.e.,
\begin{align}
    s_{\textrm{pow2}}=2^{\left\lceil\textrm{log}_{2}s\right\rceil}.
\end{align}

\subsubsection{Per-tensor scaling of activations}\label{subsec:act_scale_per_tensor}
The per-tensor activation scale is determined as follows:
\begin{subequations}
\begin{align}
    \sx =& \ \frac{\rx}{\beta r_q}\\
    \vsx =& \ \sx \mathbf{1}_{N_l}\\
    \vsc =& \ \mathbf{1}_{C_l}
\end{align}
\end{subequations}
where $\beta$ is a backoff factor that converts the measured maximal absolute input value $r_x$ into $\beta r_q$ (with $r_q$ representing the maximal quantized value), thereby providing additional headroom for representing larger input values.

The scaled activation is then computed by
\begin{align}
    \mathbf{X}_{l,s}=&\ s_{x}^{-1}\mathbf{X}_{l}.
\end{align}

\subsubsection{Per-sample scaling of activations}\label{subsec:act_scale_per_sample}
Here the scale is dynamically determined as:
\begin{subequations}
\begin{align}
\mathbf{s}_{x} =& \ \frac{\mathbf{r}_{x-}}{\beta r_q}\\
\vsc =& \ \mathbf{1}_{C_l},
\end{align}
\end{subequations}
where we omitted the batch index for brevity.
\subsubsection{Per-tensor scaling based on maximum absolute statistics of weights}\label{subsec:weight_scale_per_tensor}
For this method, the weight scales are determined by:
\begin{subequations}
\begin{align}
\sw =& \ \frac{\rw}{r_q}\\
\vsw =& \ \sw \mathbf{1}_{C_{l+1}}\\
\vsc =& \ \mathbf{1}_{C_l}
\end{align}
\end{subequations}
and the calculation of the scaled weight in (\ref{eq:vWTs}) can be simplified to:
\begin{align}
\mathbf{W}_s^{T}=s_w^{-1}\vWT.
\end{align}

\subsubsection{Per-output-channel scaling based on maximum absolute statistics of weights}\label{subsec:weight_scale_per_output_channel}
Here, the weight scales are determined by:
\begin{subequations}
\begin{align}
\vsw =& \ \frac{\mathbf{r}_{w-}}{r_q}\\
\vsc =& \ \mathbf{1}_{C_l}
\end{align}
\end{subequations}
and the calculation of the scaled weight in (\ref{eq:vWTs}) is:
\begin{align}
\mathbf{W}_s^{T}=\vWT\viSw.
\end{align}

\subsubsection{Weight quantization error minimization using per-tensor scaling}\label{subsec:weight_opt_pts}
For this method, the weight scale is determined by:
\begin{subequations}
\begin{align}
s_{w} =&\  \textrm{argmin}_{s\in\mathcal{S}}\|\vWT-sQ\left(s^{-1}\vWT\right)\|^2\\
\vsw =& \ \sw \mathbf{1}_{C_{l+1}}\\
\vsc =& \ \mathbf{1}_{C_l}
\end{align}
\end{subequations}
and the calculation of the scaled weight in (\ref{eq:vWTs}) is simplified to:
\begin{align}
\mathbf{W}_s^{T}=s_w^{-1}\vWT.
\end{align}
The set of scales that are considered in the minimization is denoted $\mathcal{S}$ and can contain arbitrary scales, power-of-2 scales, or hardware-accelerated scales.

\subsubsection{Weight quantization error minimization using per-output-channel scaling}\label{subsec:weight_opt_pcs}
Similarly to Sec.~\ref{subsec:weight_scale_per_tensor} the weight scales are determined per-output-channel by:
\begin{subequations}
\begin{align}
s_{w,k} =& \ \textrm{argmin}_{s\in\mathcal{S}}\|\vWT\vek-sQ\left(s^{-1}\vWT\vek\right)\|^2\\
\vsc =& \ \mathbf{1}_{C_l}
\end{align}
\end{subequations}
with $\mathbf{e}_k\triangleq\left[\mathbf{0}_{1\times k-1}, 1, \mathbf{0}_{1\times K-k}\right]^{T}$ being a vector that is used to extract the $k$-th column of $\vWT$ and the calculation of the scaled weight in (\ref{eq:vWTs}) is:
\begin{align}
\mathbf{W}_s^{T}=\vWT\viSw.
\end{align}

\subsubsection{SmoothQuant: per-channel scaling of activations and weights}\label{subsec:scale_smooth_quant}
In the SmoothQuant method \citep{xiao2024smoothquantaccurateefficientposttraining}, the scales of the common dimension in the matrix-multiplication are statically determined corresponding to the measured statistics of the activations (per-channel) and weights (per-input-channel):
\begin{subequations}
\begin{align}
\vsc =& \left[\frac{r_{x|,1}^{\alpha}}{r_{w|,1}^{1-\alpha}},\ldots,\frac{r_{x|,C_l}^{\alpha}}{r_{w|,C_l}^{1-\alpha}}\right]^{T}\\
s_x=&\ \frac{\max_{\dot{c}}r_{x|,\dot{c}}/{s_{c,\dot{c}}}}{\beta r_q}
\end{align}
\end{subequations}
with $\alpha$ being a parameter in the range $\left[0, 1\right]$ smoothly controlling the trade-off of selecting the per-channel scales that match the activation or weight statistics. The scaled activation (\ref{eq:vXs}) is thereby given as:
\begin{align}
\mathbf{X}_{l,s}=s_x^{-1}\mathbf{X}_{l}\viSc.
\end{align}
For calculating the weight scales we define the per-input-channel scaled weights:
\begin{align}
\bar{\mathbf{W}}^{T}=&\ \vSc\mathbf{W}^{T}.
\end{align}
The final scaled weights can be determined by additionally considering the updated per-output-channel statistics (similarly to (\ref{eq:w_maxabs_poc}):
\begin{subequations}
\begin{align}
\bar{\mathbf{r}}_{w-} =& \left[\max_{c}  |\bar{W}_{1c}|, \ldots,  \max_{c}  |\bar{W}_{C_{l+1}c}|\right]^{T}\label{eq:tw_maxabs_poc}\\
\vtsw=&\ \frac{\bar{\mathbf{r}}_{w-}}{r_q}\\
\mathbf{W}_s^{T}=&\ \vSc\vWT\vitSw
\end{align}
\end{subequations}
or updated per-tensor statistics (similarly to (\ref{eq:w_maxabs_pt}):
\begin{subequations}
\begin{align}
\bar{r}_{w} =& \max_{kc}  |\bar{W}_{kc}| \label{eq:tw_maxabs_pic}\\
\tsw =& \ \frac{\bar{r}_{w}}{r_q}\\
\mathbf{W}_s^{T}=&\ \bar{s}_w^{-1}\vSc\vWT
\end{align}
\end{subequations}


\subsection{Quantization procedure}\label{subsec:recipe}
Model quantization is a process that balances model accuracy with enhanced throughput. The procedure for quantizing a model is as follows:
\begin{enumerate}
    \item Establish an accuracy metric by selecting an evaluation dataset that is sufficiently large to reduce measurement noise while remaining computationally feasible. Also, define an acceptable accuracy degradation threshold, typically $-1\%$ or $-0.1\%$ relative to the high-precision accuracy, and determine a throughput metric.
    \item Measure the baseline accuracy and throughput of the model using high-precision computation.
    \item Perform a calibration process to compute per-tensor and per-channel statistics, utilizing a smaller calibration dataset that is preferably different from the evaluation dataset.
    \item Quantize all linear operations, i.e., matrix multiplications, and evaluate the various scaling methods. In many cases, simpler methods are prioritized as they typically have higher throughput.
    \item Consider omitting the quantization of the first and last linear layers, as these often have a greater impact on accuracy. For language models, this corresponds to the lm-head and embedding layers.
    \item Select the scaling method that meets the defined accuracy degradation threshold while achieving the highest throughput.
\end{enumerate}

%% file: 04_model_evaluations.tex
\section{Model evaluations}
In this section, we analyze the impact of different quantization schemes across model generations and sizes on a range of downstream tasks. We first analyze the throughput of FP8 matrix multiplication using different configurations, showing how different configurations can affect the overall throughput. We then evaluate models from both the Llama2, Llama3, and Mistral families, ranging from 7B to 70B parameters, focusing on three key aspects: model scale effects, task-specific impacts, and quantization method effectiveness.

\begin{table}[ht]
\centering
\begin{tabular}{@{}ccccccc@{}}
\toprule
\multicolumn{1}{c}{M} &
  \multicolumn{1}{c}{K} &
  \multicolumn{1}{c}{N} &
  \multicolumn{1}{c}{Per-Tensor} &
  \multicolumn{1}{c}{HW Accelerated} &
  \multicolumn{1}{c}{Mean TFLOPS} &
  \multicolumn{1}{c}{MFU (\%)} \\ \midrule
4096 & 4096 & 4096 & True  & True  & 803.8 & 92.9\% \\
4096 & 4096 & 4096 & True  & False & 771.4 & 89.2\% \\
4096 & 4096 & 4096 & False & False & 746.5 & 86.3\% \\ \hline
6144 & 6144 & 6144 & True  & True  & 849.1 & 98.2\% \\
6144 & 6144 & 6144 & True  & False & 837.5 & 96.8\% \\
6144 & 6144 & 6144 & False & False & 831.5 & 96.1\% \\ \hline
8192 & 8192 & 8192 & True  & True  & 851.2 & 98.4\% \\
8192 & 8192 & 8192 & True  & False & 800.8 & 92.6\% \\
8192 & 8192 & 8192 & False & False & 760.4 & 87.9\% \\ \bottomrule
\end{tabular}
\vskip 0.1in
\caption{Throughput measurements for scaled FP8 matrix multiplication of shape $(M\times K)\times (K\times N)$ in Gaudi 2. Two FP8 matrices are multiplied to produce a BF16 output matrix. We find that GEMM throughput is compute-bound for the product of matrices larger than $4096\times 4096$, with larger matrices reaching over 98\% MFU when using hardware accelerated per-tensor scaling. The peak scaled FP8 dense GEMM throughput is 865 TFLOPS. See \url{https://github.com/NAVER-INTEL-Co-Lab/gaudi-perf} for the throughput measurement code.}
\label{tab:fp8-gemm}
\end{table}

\subsection{Experimental results}
\subsubsection{Experimental setup}
All experiments were conducted on a Gaudi 2 accelerator using BF16 precision for our baseline measurements. The quantization experiments were implemented using the Intel Neural Compressor (INC) framework\footnote{\url{https://github.com/intel/neural-compressor}}, an open-source toolkit for neural network compression. Our quantization methodology applies per-tensor scaling for activations as formulated in Sec.~\ref{subsec:act_scale_per_tensor}, across all experiments, while varying between per-tensor and per-channel scaling methods for weights, as formulated in Sec.~ \ref{subsec:weight_scale_per_tensor} and Sec.~\ref{subsec:weight_scale_per_output_channel}. The \emph{Unit scale} method denotes setting the scale factors to $1$, regardless of the corresponding tensor statistics.

We evaluated model accuracy across three distinct categories: WikiText-2 \citep{merity2016pointer}) for perplexity measurement, a comprehensive common sense reasoning suite, and MMLU \citep{hendryckstest2021} for broad knowledge assessment. 
The common sense reasoning evaluation is comprised of: HellaSwag \citep{zellers2019hellaswag}, LAMBADA \citep{radford2019language}, BoolQ \citep{clark2019boolq}, ARC Easy \citep{clark2018thinksolvedquestionanswering}, PIQA \citep{Bisk2020}, Winogrande \citep{ai2:winogrande}, ARC Challenge \citep{clark2018thinksolvedquestionanswering}, and OpenBookQA \citep{OpenBookQA2018}. To measure the activation ranges for calculating the offline activation scales, we used the WebQuestions (WebQs) dataset \citep{berant-etal-2013-semantic} as a calibration set. Results are reported as the average normalized accuracy across all tasks. All datasets were evaluated using the LM Evaluation Harness framework \citep{eval-harness} in a zero-shot setting to ensure consistent comparison across different experiments.

\begin{table}[ht]
    \begin{tabular}{l|l|cc|cc|cc}
        \hline
        Model & Configuration & \multicolumn{2}{c|}{PPL - WikiText2} & \multicolumn{2}{c|}{Common sense} & \multicolumn{2}{c}{MMLU} \\
        & & Acc~$\downarrow$ & $\Delta$ (\%)~$\downarrow$ & Acc~$\uparrow$ & $\Delta$ (\%)~$\uparrow$ & Acc~$\uparrow$ & $\Delta$ (\%)~$\uparrow$ \\
        \hline
        \multirow{5}{*}{Llama2-7B} & BF16 Reference & 13.066 & -- & 67.388 & -- & 43.085 & -- \\
        & Unit Scale & 14.143 & +8.24 & 67.102 & -0.42 & 42.483 & -1.40 \\
        & Per Tensor Scaling & 13.485 & +3.20 & 67.105 & -0.42 & 40.403 & -6.23 \\
        & Per Channel Scaling & 13.477 & +3.15 & 67.307 & -0.12 & 40.377 & -6.29 \\
        \hline
        \multirow{5}{*}{Llama2-13B} & BF16 Reference & 11.465 & -- & 70.020 & -- & 54.145 & -- \\
        & Unit Scale & 11.738 & +2.38 & 70.108 & +0.13 & 53.532 & -1.13 \\
        & Per Tensor Scaling & 11.664 & +1.74 & 70.166 & +0.21 & 53.344 & -1.48 \\
        & Per Channel Scaling & 11.669 & +1.78 & 70.161 & +0.20 & 53.653 & -0.91 \\
        \hline
        \multirow{5}{*}{Llama2-70B} & BF16 Reference & 7.131 & -- & 74.652 & -- & 67.648 & -- \\
        & Unit Scale & 7.797 & +9.34 & 73.761 & -1.19 & 65.323 & -3.44 \\
        & Per Tensor Scaling & 7.279 & +2.08 & 74.336 & -0.42 & 67.504 & -0.21 \\
        & Per Channel Scaling & 7.279 & +2.07 & 74.297 & -0.48 & 67.293 & -0.53 \\
        \hline
    \end{tabular}
    \vskip 0.1in
    \caption{Llama2 model accuracy for various quantization methods.}
\end{table}

\begin{table}[ht]
    \begin{tabular}{l|l|cc|cc|cc}
        \hline
        Model & Configuration & \multicolumn{2}{c|}{PPL - WikiText2} & \multicolumn{2}{c|}{Common sense} & \multicolumn{2}{c}{MMLU} \\
        & & Acc~$\downarrow$ & $\Delta$ (\%)~$\downarrow$ & Acc~$\uparrow$ & $\Delta$ (\%)~$\uparrow$ & Acc~$\uparrow$ & $\Delta$ (\%)~$\uparrow$ \\
        \hline
        \multirow{5}{*}{Llama3-8B} & BF16 Reference & 11.069 & -- & 71.012 & -- & 65.228 & -- \\
        & Unit Scale & 11.797 & +6.58 & 70.338 & -0.95 & 63.103 & -3.26 \\
        & Per Tensor Scaling & 11.412 & +3.10 & 70.671 & -0.48 & 63.894 & -2.05 \\
        & Per Channel Scaling & 11.417 & +3.14 & 70.784 & -0.32 & 64.040 & -1.82 \\
        \hline
        \multirow{5}{*}{Llama3-70B} & BF16 Reference & 5.004 & -- & 75.751 & -- & 78.105 & -- \\
        & Unit Scale & 5.381 & +7.52 & 75.078 & -0.89 & 77.300 & -1.03 \\
        & Per Tensor Scaling & 5.176 & +3.43 & 75.582 & -0.22 & 78.252 & +0.19 \\
        & Per Channel Scaling & 5.180 & +3.52 & 75.456 & -0.39 & 77.820 & -0.37 \\
        \hline
    \end{tabular}
    \vskip 0.1in
    \caption{Llama3 model accuracy for various quantization methods.}
\end{table}
\begin{table}[ht]
    \begin{tabular}{l|l|cc|cc|cc}
        \hline
        Model & Configuration & \multicolumn{2}{c|}{PPL - WikiText2} & \multicolumn{2}{c|}{Common sense} & \multicolumn{2}{c}{MMLU} \\
        & & Acc~$\downarrow$ & $\Delta$ (\%)~$\downarrow$ & Acc~$\uparrow$ & $\Delta$ (\%)~$\uparrow$ & Acc~$\uparrow$ & $\Delta$ (\%)~$\uparrow$ \\
        \hline
        \multirow{5}{*}{Mistral-7B} & 
        BF16 Reference & 12.463 & -- & 72.239 & -- & 62.668 & -- \\
        & 
        Unit Scale & 29.45 & 136.3 & 39.667 & -45.09 & 45.584 & -27.26 \\
        & 
        Per Tensor Scaling & 13.066 & 4.84 & 72.114 & -0.17 & 60.446 & -3.55 \\
        & 
        Per Channel Scaling & 13.063 & 4.81 & 71.980 & -0.36 & 60.143 & -4.03 \\
        \hline
        \multirow{5}{*}{Mixtral-8x7B} & 
        BF16 Reference & 11.344 & -- & 72.829 & -- & 65.184 & -- \\
        & 
        Unit Scale & 96.654 & 725 & 57.381 & -21.21 & 50.832 & -22.02 \\
        & 
        Per Tensor Scaling & 11.472 & 1.13 & 73.176 & +0.48 & 64.858 & -0.50 \\
        & 
        Per Channel Scaling & 11.464 & 1.06 & 72.820 & -0.01 & 64.764 & -0.64 \\
        \hline
    \end{tabular}
    \vskip 0.1in
    \caption{Mistral model accuracy for various quantization methods.}
\end{table}





\subsection{Analysis of quantization effects}

\subsubsection{Impact of model scale}
Our experiments reveal that larger models exhibit enhanced robustness to quantization across a broad spectrum of tasks. This finding suggests that larger models likely incorporate intrinsic redundancies which serve to better preserve accuracy when precision is reduced.

\subsubsection{Task dependency}
The influence of quantization is highly task-dependent. Specifically, tasks that rely heavily on the retrieval of world knowledge, e.g., MMLU, are more susceptible to quantization effects compared to tasks that primarily engage reasoning processes. In contrast, common sense reasoning tasks designed to evaluate logical inference demonstrate remarkable robustness, with accuracy degradation typically remaining below 1\% across models.

These observations imply that quantization impacts a model's capacity to retrieve stored knowledge versus its reasoning capacity differently. This effect is particularly pronounced in smaller models and tends to diminish as model scale increases, corroborating our earlier findings.

\subsubsection{Comparison of quantization methods}
Our comparative analysis indicates distinct patterns among various quantization methods across model scales and generations. Notably, unit scale quantization consistently leads to the greatest reduction in accuracy. Conversely, the static scaling techniques evaluated, both per-tensor and per-channel scaling, offer superior preservation of accuracy. Furthermore, among these methods, per-channel scaling offers a slight advantage over per-tensor scaling, particularly in the context of smaller models.

\subsubsection{Throughput measurements}

\begin{table}[ht]
\centering
\begin{tabular}{@{}rcc@{}}
\toprule
Input Length & Mean TFLOPS & MFU (\%) \\ \midrule
1024   & 649.1       & 75.4\%       \\
2048   & 671.0       & 77.6\%       \\
4096   & 602.8       & 69.7\%       \\
8192   & 513.7       & 59.4\%       \\
16384  & 390.1       & 45.1\%       \\ \bottomrule
\end{tabular}
\vskip 0.1in
\caption{Llama v3.1 70B prefill throughputs for various sequence lengths on a single Gaudi 2. We use hardware accelerated static per-tensor quantization. The MFU values are understated because the attention operation and LM head were excluded from FP8 quantization. Peak FP8 throughput is 865 TFLOPS.}
\end{table}

\begin{table}[ht]
\centering
\begin{tabular}{@{}rrrrrr@{}}
\toprule
\textbf{}                      & \multicolumn{5}{c}{Sequence Length} \\ \midrule
\multicolumn{1}{c}{Batch size} & 512   & 1024  & 2048  & 4096 & 8192 \\ \midrule
8   & 32.8  & 32.4  & 30.8  & 30.2 & 23.4 \\
16  & 63.2  & 61.5  & 55.8  & 51.4 & 39.6 \\
32  & 120.1 & 112.0 & 94.1  & 79.5 & OOM  \\
64  & 224.1 & 198.8 & 152.3 & OOM  & OOM  \\
128 & 387.1 & 312.8 & OOM   & OOM  & OOM  \\ \bottomrule
\end{tabular}
\vskip 0.1in
\caption{Decode throughput in TFLOPS for Llama v3.1 70B models using HW acceleration with per-tensor scaling on a single Gaudi 2. Throughputs are obtained from time measurements of 256 decode steps before the target length.}
\end{table}

Using the method proposed in \citet{kim2025investigationfp8acceleratorsllm}, we measure the throughput of Llama v3.1 70B models on a single Gaudi 2 for both the prefill and decode phases. We use FP8 quantization only on the linear layers of the model, excluding the LM head. Measurements are in model FLOPS using the formula used in \citet{kim2025investigationfp8acceleratorsllm}, which exclude FLOPs from the attention mask.

For both prefill and decode phases, we find that throughput is best at shorter sequence lengths. This is expected, as FP8 is not applied to the attention computation. However, even for 8096 long sequences, we can see that FP8 improves prefill throughput to levels above the peak BF16 GEMM throughput. Also, thanks to the memory gain, we can measure Llama 70B on a single Gaudi 2, which would not be possible with BF16. This allows us to measure throughputs without being affected by tensor parallel overheads.

%% file: 05_discussion.tex
\section{Conclusion}
In this report, we have detailed the mechanism behind scaled FP8 matrix multiplication in Intel Gaudi accelerators. By providing a reference for how the scaled FP8 matrix-multiplication works in Gaudi accelerators, we aim to help the community in leveraging the FP8 data type for their use cases. Furthermore, by providing empirical results for throughput and accuracy, we provide a point of comparison for those seeking to accelerate their workloads with FP8.